\documentstyle[12pt,epsfig,axodraw,a4]{article} 
\def\bild#1#2{    
        \vspace*{-5mm}
        \begin{center}
        \begin{math}
        \epsfxsize#2cm
        \epsffile{#1}
        \end{math}
        \end{center}  }
\newcommand{\vs}{\vspace{-0.25cm}}
\begin{document} 
\begin{center}
\large{\bf Pion-photon exchange nucleon-nucleon potentials}

\medskip

N. Kaiser\\

\smallskip

{\small Physik Department T39, Technische Universit\"{a}t M\"{u}nchen,
    D-85747 Garching, Germany}

\end{center}

\bigskip

\begin{abstract}
We calculate in chiral perturbation theory the dominant next-to-leading order 
correction to the $\pi\gamma$-exchange NN-potential proportional to the large
isovector magnetic moment $\kappa_v = 4.7$ of the nucleon. The corresponding 
spin-spin and tensor potentials $\widetilde V_{S,T}(r)$ in coordinate space 
have a very simple analytical form. At long distances $r \simeq 2$\,fm these 
potentials  are of similar size (but opposite in sign) as the leading order 
$\pi\gamma$-exchange potentials. We consider also effects from virtual 
$\Delta$-isobar excitation as well as other isospin-breaking contributions to 
the $2\pi$-exchange NN-potential induced by additional one-photon exchange.    
\end{abstract}


\bigskip

To be published in {\it The Physical Review C (2006), Brief Reports}

\bigskip

\vspace{1.0cm}
Isospin violation in the nuclear force is a subject of current interest. 
Charge-independence breaking (i.e. the difference between the total isospin 
$I=1$ $pn$-scattering and $nn$- or $pp$-scattering) is large and well 
established. On the other hand, charge-symmetry breaking (i.e. the difference
between  $nn$- and $pp$-scattering after removal the long-range
electromagnetic forces) is smaller and fairly well established. These 
isospin-violating contributions to the nuclear force play also an important
role for explaining the $764\,$keV binding energy difference of $^3$He and
triton \cite{friar1,brand,wu}. The bulk of it, namely $648\,$keV, can already
be understood in terms of the Coulomb interaction \cite{friar1}.     

The isospin-violating NN-interaction of longest range is generated by the 
simultaneous exchange of a pion and a photon between the two nucleons. Since
the photon is massless the $\pi\gamma$-exchange interaction is of nominal
one-pion range, $m_\pi^{-1} = 1.41\,$fm. After earlier attempts 
made in refs.\cite{leung,friar} the complete leading order 
$\pi\gamma$-exchange NN-potential has been calculated in the systematic 
framework of chiral perturbation theory in ref.\cite{kolck}. A crucial feature
of that calculation has been to guarantee the gauge-invariance of the result 
by considering the full set of all 19 possible Feynman diagrams. The resulting
expression for the complete leading order $\pi\gamma$-exchange potential in 
momentum space turned out to be surprisingly simple, see Eq.(1) below. 
However, due to its intrinsic smallness (a few permille of the $1\pi$-exchange 
interaction) the inclusion of this new isospin-breaking interaction had 
negligible effects on the $^1S_0$ low-energy parameters and it lead only to a 
tiny improvement in the fits of the NN-scattering data \cite{kolck}. 
Nevertheless it is important to know as accurately as possible the size of 
these well-defined long-range components before one introduces (adjustable) 
short-range isospin-violating terms.  

The purpose of the present short paper is to calculate next-to-leading order
corrections to the $\pi\gamma$-exchange NN-potential in the systematic 
framework of chiral perturbation theory. These corrections arise either as 
relativistic $1/M$-corrections (with $M=939\,$MeV the average nucleon mass) to the
static result of ref.\cite{kolck} or they are generated by new interaction
vertices from the next-to-leading order chiral Lagrangian ${\cal L}^{(2)}_{\pi
N}$ \cite{review}. Experience with the isospin-conserving NN-potential has
shown that the next-to-leading order corrections are dominated by the  
contributions proportional to the large low-energy constants, in that case 
$c_{1,3,4}$ \cite{nnpot}. For the photon-nucleon coupling which is of
relevance in the present work, one readily identifies the isovector magnetic 
moment $\kappa_v =4.7$ of the nucleon as an outstandingly large low-energy 
parameter. Therefore we will focus here on this particular contribution to the 
$\pi\gamma$-exchange NN-potential. Effects from virtual $\Delta(1232)$-isobar
excitation which involve the equally strong $\Delta\to N \gamma$ transition
magnetic moment $\kappa^* \simeq 4.9$ will also be considered. 

Let us start with reanalyzing the leading order $\pi\gamma$-exchange 
NN-potential of ref.\cite{kolck}. The corresponding T-matrix in momentum space
reads:
\begin{equation} T_{\pi \gamma}^{(lo)}= {\alpha g_A^2 \over 8\pi f_\pi^2}
(\vec\tau_1 \cdot\vec\tau_2-\tau_1^3 \tau_2^3)\,\vec \sigma_1 \cdot\vec q \,\, 
\vec \sigma_2 \cdot\vec q\,\, \bigg\{ {1\over q^2}-{(m_\pi^2- q^2)^2 \over q^4 
(m_\pi^2+ q^2)} \ln\bigg(1+ {q^2\over m_\pi^2}\bigg) \bigg\}\,, \end{equation} 
with $\alpha = 1/137.036$ the fine structure constant, $g_A =g_{\pi N} f_\pi/M
= 1.3$ the nucleon axial vector coupling constant, $f_\pi = 92.4$\,MeV the 
pion decay constant, and $m_\pi = 139.57 $\,MeV the charged pion mass. 
Furthermore, $\vec q$ denotes the momentum transfer between both nucleons, and
$\vec \sigma_{1,2}$ and $\vec \tau_{1,2}$ are the usual spin- and isospin 
operators of the two nucleons. The Fourier transformation, $-(2\pi)^{-3} \int
d^3q  \exp(i \vec q \cdot \vec r\, )\dots $, of Eq.(1) to coordinate space
yields a local potential with spin-spin and tensor components: 
\begin{equation} \Big\{ \widetilde V_S(r) \, \vec \sigma_1 \cdot \vec \sigma_2 
+ \widetilde V_T(r) \, (3 \vec \sigma_1 \cdot \hat r\, \vec\sigma_2\cdot \hat 
r - \vec \sigma_1 \cdot \vec  \sigma_2) \Big\}\, {1\over 2}(\vec\tau_1 \cdot
\vec\tau_2-\tau_1^3\tau_2^3)\,. \end{equation}
The isospin factor $(\vec\tau_1 \cdot\vec\tau_2-\tau_1^3\tau_2^3)/2$ has been
chosen such that is gives $1$ for elastic $pn\to np$ scattering.
The leading order $\pi\gamma$-exchange spin-spin potential reads:
\begin{equation} \widetilde V^{(lo)}_S(r) = { \alpha g_A^2 m_\pi^2 \over (4\pi
f_\pi)^2} \,{2e^{-x} \over 3r} \,\bigg\{ 2 \ln{x\over 2} +2 \gamma_E 
-{1\over x} - {1\over x^2} + \widetilde E(x)-2 \widetilde  E(2x) \bigg\} \,,
\end{equation}
with $x = m_\pi r$ and $\widetilde E(x) = \int_0^\infty d\zeta \, e^{-\zeta}
(\zeta+x)^{-1}$ the modified exponential integral function. $\gamma_E \simeq 
0.5772$ is the Euler-Mascheroni number. The associated tensor potential has a
similar form: 
\begin{eqnarray} \widetilde V^{(lo)}_T(r)&=& {\alpha g_A^2 \over (4\pi f_\pi
)^2}\, { e^{-x} \over 3r^3}\, \bigg\{x-5+4(3+3x+x^2) \bigg(\ln{x\over 2} +
\gamma_E \bigg) \nonumber \\ && +( 18 -x^2) \widetilde E(x)+4(3x-3-x^2)
\widetilde  E(2x)\bigg\}  \,.\end{eqnarray}
In the first and second row of Table\,I we present some numerical values of 
these leading order $\pi\gamma$-exchange potentials for distances $1\,{\rm fm}
\leq r \leq 3 \,{\rm fm}$. Somewhat as a surprise, one observes a
non-monotonic dependence on the nucleon distance $r$. The spin-spin potential
$\widetilde V^{(lo)}_S(r)$ passes through zero at $r = 2.53\,$fm and it has a
maximum at $r =3.34\,$fm. For the tensor potential $\widetilde V^{(lo)}_T(r)$
these points are shifted inward and approximately reduced by a factor 1.9. It 
passes through zero at $r = 1.31\,$fm and it has its maximum value of about
$8.8$\,keV at $r = 1.78\,$fm.

\begin{table}[hbt]
\begin{center}
\begin{tabular}{|c|ccccccccccc|}
    \hline

    $r$~[fm]& 1.0 & 1.2 & 1.4 & 1.6 & 1.8 & 2.0 & 2.2 & 2.4 & 2.6& 2.8 & 3.0\\
   \hline
$\widetilde V_S^{(lo)}$ &--52.3 &--27.4 & --15.0 &--8.30 & --4.54 &--2.35
   & --1.07 &--0.31 &0.13 & 0.37 & 0.50 \\
    
$\widetilde V_T^{(lo)}$ & --48.3 & --9.26 & 3.93 & 8.06 & 8.79 & 8.26 & 
7.31 & 6.29 & 5.34 & 4.49 & 3.76 \\

$\widetilde V_S^{(\kappa_v)}$ & 61.2 & 27.7 & 14.0 & 7.63 & 4.41 & 2.67 & 1.67
& 1.08 &0.72 & 0.49 & 0.34  \\ 
    
$\widetilde V_T^{(\kappa_v)}$& --230 & --100 & --49.1 &--26.0 &--14.6 & 
--8.64 & --5.31 & --3.37 & --2.19 & --1.46  & --0.99\\
$\widetilde V_S^{(\kappa^*)}$ & 14.3 & 5.87 & 2.70 & 1.35 & 0.72 & 0.41 & 
0.24 & 0.15 & 0.09  & 0.06 & 0.04 \\
    
$\widetilde V_T^{(\kappa^*)}$& --36.4 & --14.3 & --6.32 &--3.07
    &--1.60 & --0.88& --0.51 & --0.30 & --0.19 & --0.12  & --0.08\\ 
    \hline
\end{tabular}
\end{center}
{\it Table I: Pion-photon exchange pn-potentials in units of keV versus the 
nucleon distance $r$. The spin-spin and tensor potentials $\widetilde 
V_{S,T}^{(lo)}$ correspond to the leading order in the chiral expansion. The
next-to-leading order corrections $\widetilde V_{S,T}^{(\kappa_v)}$ are  
proportional to the large isovector magnetic moment $\kappa_v= 4.7$ of the 
nucleon, and $\widetilde V_{S,T}^{(\kappa^*)}$ arise from the magnetic 
$\Delta \to  N\gamma$ transition.}
\end{table}

\bigskip

In order to arrive at the analytical results Eqs.(1,3,4) one can actually 
circumvent the complete evaluation of all contributing one-loop diagrams. It
is sufficient to calculate their spectral function or imaginary part using
the Cutkosky cutting rule. The pertinent two-body phase space integral is 
most conveniently performed in the $\pi\gamma$ center-of-mass frame where it
becomes proportional to a simple angular integral: $(\mu^2-m_\pi^2)/(32\pi 
\mu^2) \int_{-1}^1dz\dots $, with $\mu \geq m_\pi$ the $\pi\gamma$ invariant 
mass. Gauge invariance can be controlled by working with the (generalized) 
photon propagator $(-g_{\mu\nu}+\xi\, k_\mu k_\nu/k^2)/k^2$ through the
$\xi$-independence of the total spectral function. Using these techniques we
obtain from the leading order one-loop $\pi\gamma$-exchange diagrams of
ref.\cite{kolck}: 
\begin{equation} {\rm Im}\,T_{\pi\gamma}^{(lo)} ={\alpha g_A^2 \over 8 f_\pi^2}
(\vec\tau_1 \cdot\vec\tau_2-\tau_1^3 \tau_2^3)\,\vec \sigma_1 \cdot\vec q \,\, 
\vec \sigma_2 \cdot\vec q\,\, {(\mu^2+m_\pi^2)^2 \over \mu^4 (m_\pi^2-\mu^2)} 
\,. \end{equation}      
The notation Im\,$T_{\pi\gamma}$ is meant here such that one is taking the
imaginary part of the amplitude standing to the right of the spin- and isospin 
factors.

\bigskip
\bild{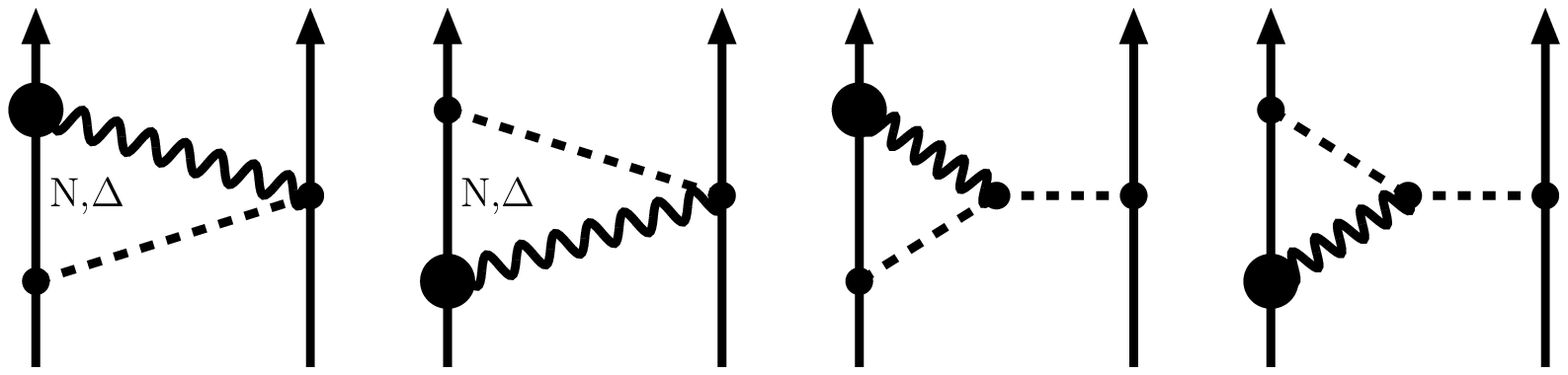}{16}
\vskip 0.3cm
{\it Fig.\,1: Pion-photon exchange diagrams generating a non-vanishing
imaginary part. The heavy dot symbolizes the magnetic coupling of the photon
to the nucleon, or the magnetic $\Delta \to N \gamma$ transition. Diagrams for 
which the role of both nucleons is interchanged are not shown. These lead
effectively to a doubling of the NN-potential.} 
\bigskip

Now we turn to the dominant next-to-leading correction to the
$\pi\gamma$-exchange NN-potential proportional to the large isovector magnetic
moment $\kappa_v = 4.7$. The relevant one-loop diagrams with a nucleon in the
intermediate state are shown in Fig.\,1. The pertinent Feynman rules can be
found in appendix A of ref.\cite{review}. From the calculated spectral
function, $(\mu^2-m_\pi^2)/\mu^5$ times a polynomial in $\mu^2$ and $m_\pi^2$, 
we can derive (via a once-subtracted dispersion relation) the following 
expression for the T-matrix in momentum space:  
\begin{eqnarray} T_{\pi \gamma}^{(\kappa_v)}&=& {\alpha g_A^2\kappa_v\over 64M
f_\pi^2 q^3} (\vec\tau_1 \cdot\vec\tau_2-\tau_1^3 \tau_2^3)\bigg\{\vec\sigma_1 
\cdot \vec \sigma_2 \bigg[ (m_\pi^2 +q^2)(3q^2-m_\pi^2) \arctan{q \over m_\pi}
+ m_\pi^3 q \bigg] \nonumber \\ && + {1\over q^2} \,\vec \sigma_1 \cdot \vec q
\,\, \vec \sigma_2 \cdot\vec q\,\bigg[ (m_\pi^2 +q^2)(3m_\pi^2-5q^2) \arctan{q 
\over m_\pi} + 3m_\pi q(q^2-m_\pi^2)\bigg] \bigg\}\,. \end{eqnarray}
As a side remark we note that the contribution proportional to the isoscalar 
magnetic moment $\kappa_s = 0.88$ vanishes identically. The reason for this
feature are the vanishing angular integrals: $-\!\!\!\!\!\int_{-1}^1 dz\,z^{-1}
= 0 = \int_{-1}^1 dz\, z$. Fourier transformation of Eq.(6) to coordinate
space yields spin-spin and tensor potentials of the following simple
analytical form:   
\begin{equation} \widetilde V^{(\kappa_v)}_S(r) = {\alpha g_A^2 \kappa_v 
\over 48 \pi M f_\pi^2}\, {e^{-m_\pi r}\over r^4 }(1+m_\pi r) \,,\end{equation}
\begin{equation} \widetilde V^{(\kappa_v)}_T(r) = -{\alpha g_A^2 \kappa_v\over 
48 \pi M  f_\pi^2}\, {e^{-m_\pi r}\over r^4 } (5+2m_\pi r) \,.\end{equation}
The more direct way to obtain these analytical expressions is to ''Laplace 
transform'' the spectral function Im$\,T_{\pi \gamma}^{(\kappa_v)}$ (for 
details see Eqs.(3,4) in ref.\cite{2looppot}). In the third and fourth row of 
Table\,I we have collected some numerical values of these novel 
isospin-violating potentials $V^{(\kappa_v)}_{S,T}(r)$. At long 
distances $r \simeq 2 \,{\rm fm}$ they are of the same size but opposite in 
sign as the leading order $\pi\gamma$-exchange potentials (compare with the 
first and second row in Table\,I). At shorter distances the tensor component 
$V^{(\kappa_v)}_T(r)$ in Eq.(8) with its higher weight factor on the singular 
$1/r^4$-term becomes in fact dominant. Note also that there is some tendency 
for cancellation in the long-range tails of the tensor potentials.

\bigskip
\bigskip

\bild{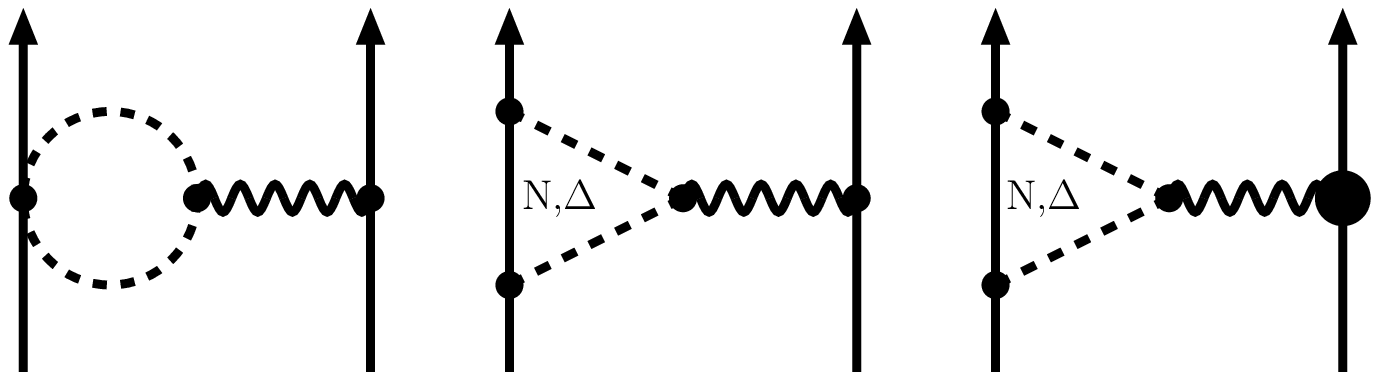}{14}
\vskip 0.3cm

{\it Fig.\,2: Isospin-breaking corrections to the two-pion exchange
NN-interaction induced by one-photon exchange. The heavy dot in the right
diagram symbolizes the magnetic coupling of the photon to the nucleon. The
$2\pi$-exchange diagrams for which the role of both nucleons is interchanged
are not shown.} 

\bigskip

As argued in ref.\cite{friar}, the $\Delta(1232)$-isobar with its relatively 
small excitation energy could also play a substantial role for the $\pi\gamma
$-exchange interaction. The transition $\Delta\to N \gamma$ is known to be 
predominantly of magnetic dipole type. In the effective chiral Lagrangian 
approach its strength is parameterized by a transition magnetic moment
$\kappa^*$. Using the empirical information \cite{pdg} about the partial decay
width:   
\begin{equation} \Gamma(\Delta^+ \to p \gamma) = { \alpha \kappa^{*2}
(M_\Delta^2-M^2)^3 \over 144M_\Delta^5M^2} (3M_\Delta^2+M^2) \simeq 0.68 \,{\rm
MeV}  \,, \end{equation} 
we can extract a value of $\kappa^* \simeq 4.9$ for the $\Delta\to N \gamma$ 
transition magnetic moment. Here, $M_\Delta = 1232$\,MeV denotes the mass of
the delta-isobar. The possible one-loop diagrams with virtual excitation of a
delta-resonance contributing to the $\pi\gamma$-exchange NN-interaction are
shown in Fig.\,1. The Feynman rules for the non-relativistic $\pi N \Delta$-
and $\gamma N \Delta$-vertices read: $(g_{\pi N \Delta}/2M) \, \vec S \cdot
\vec k\,T^a$ and $(e\kappa^*/2M) \, \vec S \cdot (\vec k \times\vec\epsilon\,) 
\, T^3$, respectively, where $\vec k$ denotes an ingoing pion or photon 
momentum. The $2\times 4$ spin- and isospin transition matrices $S^i$ and
$T^a$ satisfy the relations $S^i S^{\dagger j} = (2 \delta^{i j}- i 
\epsilon^{ijk} \sigma^k)/3$ and $T^a T^{\dagger b} = (2 \delta^{a b}- i 
\epsilon^{abc} \tau^c)/3$. Using the empirically well-satisfied relation 
$g_{\pi N \Delta} = 3 g_{\pi N}/\sqrt{2}= 3 g_A M/\sqrt{2} f_\pi$ for the $\pi
N \Delta$-coupling constant we find the following result for their total 
spectral function:         
\begin{eqnarray} {\rm Im}\,T_{\pi \gamma}^{(\kappa^*)}&=&{\alpha g_A^2 \kappa^*
\over 96 \sqrt{2} M f_\pi^2 \mu^3} (\vec\tau_1 \cdot\vec\tau_2-\tau_1^3 
\tau_2^3) \bigg\{\vec \sigma_1 \cdot \vec  \sigma_2 \bigg[ 2\mu \Delta (\mu^2 
- m_\pi^2)\nonumber \\ && + (m_\pi^4 +2m_\pi^2 \mu^2 -3\mu^4 -4\mu^2 \Delta^2)
\arctan{\mu^2 - m_\pi^2 \over 2 \mu \Delta} \bigg]\nonumber \\ &&+\vec
\sigma_1 \cdot \vec q \,\,  \vec  \sigma_2 \cdot\vec q \, \bigg[ 2\Delta
\bigg({ m_\pi^2 \over \mu }+ 3\mu \bigg) \nonumber \\ && +\bigg(2m_\pi^2 -5
\mu^2 +{3m_\pi^4 \over \mu^2} -{4\Delta^2 (3\mu^2 +m_\pi^2) \over \mu^2 -
m_\pi^2 }\bigg)\arctan{ \mu^2 - m_\pi^2 \over  2 \mu \Delta}\bigg] \bigg\}\,,
\end{eqnarray}  
where $\Delta= M_\Delta -M =293\,$MeV denotes the delta-nucleon mass
difference. Note that the spectral function of the spin-spin term
(proportional to $\vec \sigma_1 \cdot \vec  \sigma_2$) and the spectral
function of  the tensor term (proportional to $\vec \sigma_1 \cdot  \vec q \,
\,  \vec  \sigma_2 \cdot\vec q\,$) vanish  both at threshold $\mu = m_\pi$. 
From the mass-spectra given by Eq.(10) one can easily calculate the spin-spin 
and tensor potentials in coordinate space (following the decomposition in
Eq.(2)) in the form of a continuous superposition of Yukawa functions
\cite{2looppot}. The fifth and sixth row in Table\,I shows the corresponding
numerical values for the $\pi\gamma$-exchange potentials
$V^{(\kappa^*)}_{S,T}(r)$ for nucleon distances $1\,{\rm fm}\leq r \leq 3
\,{\rm fm}$. One finds that the effects from virtual $\Delta$-excitation are
about a factor 5 to 10 smaller than those generated by diagrams with only
nucleon intermediate states. About such a suppression of the $\Delta$-isobar
effects has already been speculated in the summary of ref.\cite{friar}. The
present calculation provides now a quantitative answer to this question.

Fig.\,2 shows another set of isospin-violating contributions to the $2\pi
$-exchange NN-interaction induced by an additional one-photon exchange. These
effects could alternatively be interpreted as one-pion loop contributions to 
the  electric and magnetic form factors of the nucleon which are introduced in
order to describe the electromagnetic interaction between the extended (not 
point-like) nucleons. Irrespective of their classification the magnitude of 
such isospin-breaking effects should be quantified. The first two diagrams in 
Fig.\,2 (allowing only for an intermediate nucleon state) with the photon 
coupling to the charge of the nucleon give rise to the following T-matrix:     
\begin{eqnarray} T_{\pi \gamma}^{(lo)}&=& {\alpha \over 48\pi f_\pi^2}
(\tau_1^3+ \tau_2^3+2\tau_1^3 \tau_2^3) \bigg[ 1+5g_A^2 +{4m_\pi^2 \over q^2}
(1+2g_A^2) \bigg] \nonumber \\ && \times \Bigg\{ 1 - {\sqrt{4m_\pi^2+q^2} 
\over q} \ln{q+ \sqrt{4m_\pi^2+q^2} \over 2m_\pi} \Bigg\} \,. \end{eqnarray}
The corresponding central potential in coordinate space:
\begin{eqnarray} \widetilde V^{(lo)}_C(r) &=& {\alpha \over 3(8\pi f_\pi)^2\,r}
(\tau_1^3 + \tau_2^3+2\tau_1^3 \tau_2^3) \nonumber \\ && \times 
\int_{2m_\pi}^\infty d\mu\, e^{-\mu r} 
\sqrt{\mu^2-4m_\pi^2 }\, \bigg[ {4m_\pi^2 \over \mu^2}(1+2g_A^2)- 1-5g_A^2
\bigg] \,, \end{eqnarray}
has some similarity with the Uehling potential. As the numbers in the first row
of Table\,II show it is attractive and of similar size as the leading order
$\pi\gamma$-exchange spin-spin potential (see the first row in Table\,I). We 
have evaluated here $\widetilde V^{(lo)}_C(r)$ for $pp$-scattering where the
isospin factor $\tau_1^3 + \tau_2^3+2\tau_1^3 \tau_2^3$ becomes equal to
$4$. The second diagram in Fig.\,2 with a virtual $\Delta$-excitation leads to
the spectral function: 
\begin{eqnarray} {\rm Im}\,T_{\pi \gamma}^{(\Delta)}&=& {\alpha g_A^2 \over 8 
f_\pi^2 \mu^3} (\tau_1^3+ \tau_2^3+2 \tau_1^3 \tau_2^3) \Bigg\{ \bigg(
{2m_\pi^2 \over 3} - \Delta^2 - {5 \mu^2 \over 12} \bigg) \sqrt{\mu^2 -
4m_\pi^2}  \nonumber \\ && + \Delta (\mu^2 -2m_\pi^2 +2\Delta^2)
\arctan{\sqrt{\mu^2 -  4m_\pi^2} \over 2 \Delta}  \Bigg\}\,. \end{eqnarray}
The corresponding central potential $\widetilde V^{(\Delta)}_C(r)$ (see second
row in Table\,II) comes out repulsive and it is approximately an order of 
magnitude smaller than the leading order one $\widetilde V^{(lo)}_C(r)$. The
last diagram in Fig.\,2 involves the magnetic coupling of the photon to the
nucleon. We are considering only the dominant contribution 
proportional to the isovector magnetic moment $\kappa_v = 4.7$ and find for the
corresponding one-loop T-matrix:       
\begin{equation} T_{\pi \gamma}^{(N)}= {\alpha g_A^2 \kappa_v\over 32M f_\pi^2
} \,\tau_1^3 \tau_2^3\,(\vec \sigma_1 \times\vec q \,)\cdot (\vec \sigma_2
\times \vec q\,) \,\bigg\{ {2m_\pi \over q^2}-{4m_\pi^2+ q^2 \over q^3}
\arctan{ q \over 2 m_\pi } \bigg\}\,. \end{equation} 
When translated into coordinate space one obtains a spin-spin and tensor 
potential:   
\begin{equation} \widetilde V^{(N)}_S(r) = -{\alpha g_A^2 \kappa_v \over 96 
\pi M  f_\pi^2}\,\tau_1^3 \tau_2^3\, {e^{-2m_\pi r}\over r^4 } (1+2m_\pi r)
\,,\end{equation} 
\begin{equation} \widetilde V^{(N)}_T(r) = {\alpha g_A^2 \kappa_v \over 96 \pi 
M  f_\pi^2}\,\tau_1^3 \tau_2^3\,{e^{-2m_\pi r}\over r^4 } (2+m_\pi r)
\,,\end{equation} 
of the typical two-pion range $(2m_\pi)^{-1} = 0.7\,$fm. As the numbers in the
third and fourth row of Table\,II indicate they differ from each other mainly 
in sign. The magnitude of $V^{(N)}_{S,T}(r)$ comes out substantially smaller
than that of the central potential $V^{(lo)}_C(r)$. This is to be expected
since the magnetic interaction is a higher order relativistic
$1/M$-correction. Finally, we include also a virtual $\Delta$-isobar in this 
two-pion exchange process followed by one-photon exchange. The corresponding 
spectral function:  
\begin{eqnarray} {\rm Im}\, T_{\pi \gamma}^{(\Delta)}&=& {\alpha g_A^2\kappa_v 
\over 64M f_\pi^2 \mu^3 }\,\tau_1^3 \tau_2^3\,(\vec \sigma_1\cdot\vec \sigma_2
\,\mu^2 + \vec \sigma_1 \cdot\vec q \,\,\vec \sigma_2 \cdot \vec q\,)
\nonumber \\ && \times \Bigg\{ -2\Delta \sqrt{\mu^2- 4m_\pi^2} + (\mu^2 +4
\Delta^2 -  4m_\pi^2 ) \arctan{\sqrt{\mu^2 - 4m_\pi^2} \over 2 \Delta}
\Bigg\}\,,  \end{eqnarray}
leads to the numerical values of the isospin violating spin-spin and tensor 
potentials $V^{(\Delta)}_{S,T}(r)$ presented in the fifth and sixth row of
Table\,II. These potentials are of the same sign but a factor of 5 to 10 
smaller than their counterparts $V^{(N)}_{S,T}(r)$ with pure nucleon 
intermediate states.  

\begin{table}[hbt]
\begin{center}
\begin{tabular}{|c|ccccccccc|}
    \hline

    $r$~[fm]& 1.0 & 1.2 & 1.4 & 1.6 & 1.8 & 2.0 & 2.2 & 2.4 & 2.6\\
   \hline
$\widetilde V_C^{(lo)}$ & --49.5 & --22.5 & --11.0   &--5.74 & --3.13 & --1.76
   & --1.02 & --0.61 & --0.37 \\
    
$\widetilde V_C^{(\Delta)}$ & 7.68 &2.96 & 1.26 &
0.58 & 0.28 & 0.14 & 0.08 & 0.04 & 0.02 \\

$\widetilde V_S^{(N)}$ &--21.3 & --8.66 & --3.89 & --1.88 & --0.96 & --0.51 & 
--0.28 & --0.16 & --0.09 \\ 
    
$\widetilde V_T^{(N)}$& 23.9 & 9.14 & 3.90 & 1.81 & 0.89 & 0.46 & 0.25 & 0.14 
& 0.08\\

$\widetilde V_S^{(\Delta)}$ & --4.09 & --1.50 & --0.62 & --0.28 & --0.13  &
--0.07  & --0.03 & --0.02 & --0.01 \\
    
$\widetilde V_T^{(\Delta)}$& 4.18 & 1.46 & 0.58 & 0.25 & 0.12 & 0.06 & 0.03 & 
0.02 & 0.01\\ 
\hline
$\delta \widetilde V_{2\pi}^{(cib)}$  & 108 & 55.6 & 31.1 & 18.4 & 11.4
& 7.25  & 4.74 & 3.16 & 2.14 \\
$\widetilde V_{C}^{(csb)}$  & --182 & --80.0 & --38.9 & --20.3 & --11.3
& --6.51  & --3.89 & --2.39 & --1.50 \\
$\widetilde V_{S}^{(csb)}$ & 67.3 & 29.6 & 14.4 & 7.50 & 4.11
& 2.34  & 1.38 & 0.83 & 0.51 \\
$\widetilde V_{T}^{(csb)}$  & --84.1 & --34.7 & --15.9 &
--7.86  & --4.11 & --2.26  & --1.28 & --0.75 & --0.45 \\
\hline
\end{tabular}
\end{center}
{\it Table\,II: Isospin violating contributions to the two-pion exchange
pp-potential in units of keV versus the distance $r$. The spin-spin and tensor
potentials $V_{S,T}^{(N,\Delta)}$ are proportional to the large isovector 
magnetic moment $\kappa_v = 4.7$. The charge-independence breaking potential
$\delta \widetilde V_{2\pi}^{(cib)}$ proportional to the pion mass difference
$m_{\pi^+}-m_{\pi^0} = 4.59\,$MeV is taken from ref.\cite{cib}. The 
charge-symmetry breaking potentials $\widetilde V_{C,S,T}^{(csb)}$
proportional to the neutron-proton mass difference $M_n- M_p = 1.29\,$MeV are  
taken from ref.\cite{csb}.}  
\end{table}

It is also instructive to compare our present results with previously calculated
isospin-breaking $2\pi$-exchange potentials. Taking into account the mass
difference between the charged and neutral pion, $m_{\pi^+}-m_{\pi^0}=4.59\,$MeV,
in the pion-loops ref.\cite{cib} obtained the charge-independence breaking central
potential $\delta\widetilde V_{2\pi}^{(cib)}(r)$ (for an explicit expression, see 
Eq.(11) in ref.\cite{cib}). Moreover, the neutron-proton mass difference,
$M_n - M_p = 1.29\,$MeV, in intermediate nucleon states of the pion-loops leads
to the charge-symmetry breaking potentials $\widetilde V_{C,S,T}^{(csb)}(r)$
(proportional to $\tau_1^3+\tau_2^3$) derived recently in ref.\cite{csb} (for
details see Eq.(10) therein). As one can see from the numerical values in 
Table\,II the effects of these hadron mass splittings are substantially larger 
than the one-photon exchange corrections studies here. For more extensive
recent work on isospin-violating NN-forces using the method of unitary 
transformations, see also ref.\cite{evgeni}.

In summary we calculated in this work next-to-leading order corrections to the 
$\pi\gamma$-exchange NN-potential. The dominant contribution proportional to
the large isovector magnetic moment $\kappa_v = 4.7$ turns out to be of 
similar size (but opposite in sign) as the leading order term. Effects from 
virtual $\Delta$-isobar excitation, involving the equally large $\Delta \to N
\gamma$ transition magnetic moment $\kappa^* \simeq 4.9$, are approximately
one order of magnitude smaller. Furthermore, we have also evaluated several
isospin-violating contributions to the $2\pi$-exchange NN-potential induced by
additional one-photon exchange. In most cases these turned out to be smaller 
than the $\pi\gamma$-exchange terms and the effects from pion and nucleon mass
splittings \cite{cib,csb}. The analytical expressions for the
T-matrices and coordinate space potentials derived in this work are in a form
that they can be easily implemented into NN-phase shift analyses or few-body 
calculations. Such numerical studies will reveal the role of the  long-range 
isospin-violating NN-interaction generated by pion-photon exchange.

\end{document}